\documentclass[conference,a4paper]{IEEEtran}
\IEEEoverridecommandlockouts

\usepackage{mathtools,bm}
\usepackage{tikz, pgfplots}
\usetikzlibrary{positioning, decorations.text, shapes.geometric, arrows, calc, fit, matrix, positioning, shapes, shadows, trees, mindmap, tikzmark, arrows.meta, angles, quotes, babel, spy, decorations.markings}
\usepackage{url}
\usepackage{siunitx}
\usepackage{cite}
\usepackage{amsmath,amssymb,amsfonts}
\usepackage{bbm}
\usepackage{algorithmic}
\usepackage{graphicx}
\usepackage{textcomp}
\usepackage{xcolor}
\def\BibTeX{{\rm B\kern-.05em{\sc i\kern-.025em b}\kern-.08em
    T\kern-.1667em\lower.7ex\hbox{E}\kern-.125emX}}
\definecolor{kit-green100}{rgb}{0,.59,.51}
\definecolor{kit-green70}{rgb}{.3,.71,.65}
\definecolor{kit-green50}{rgb}{.50,.79,.75}
\definecolor{kit-green30}{rgb}{.69,.87,.85}
\definecolor{kit-green15}{rgb}{.85,.93,.93}
\definecolor{KITgreen}{rgb}{0,.59,.51}

\definecolor{KITpalegreen}{RGB}{130,190,60}
\colorlet{kit-maigreen100}{KITpalegreen}
\colorlet{kit-maigreen70}{KITpalegreen!70}
\colorlet{kit-maigreen50}{KITpalegreen!50}
\colorlet{kit-maigreen30}{KITpalegreen!30}
\colorlet{kit-maigreen15}{KITpalegreen!15}

\definecolor{KITblue}{rgb}{.27,.39,.66}
\definecolor{kit-blue100}{rgb}{.27,.39,.67}
\definecolor{kit-blue70}{rgb}{.49,.57,.76}
\definecolor{kit-blue50}{rgb}{.64,.69,.83}
\definecolor{kit-blue30}{rgb}{.78,.82,.9}
\definecolor{kit-blue15}{rgb}{.89,.91,.95}

\definecolor{KITyellow}{rgb}{.98,.89,0}
\definecolor{kit-yellow100}{cmyk}{0,.05,1,0}
\definecolor{kit-yellow70}{cmyk}{0,.035,.7,0}
\definecolor{kit-yellow50}{cmyk}{0,.025,.5,0}
\definecolor{kit-yellow30}{cmyk}{0,.015,.3,0}
\definecolor{kit-yellow15}{cmyk}{0,.0075,.15,0}

\definecolor{KITorange}{rgb}{.87,.60,.10}
\definecolor{kit-orange100}{cmyk}{0,.45,1,0}
\definecolor{kit-orange70}{cmyk}{0,.315,.7,0}
\definecolor{kit-orange50}{cmyk}{0,.225,.5,0}
\definecolor{kit-orange30}{cmyk}{0,.135,.3,0}
\definecolor{kit-orange15}{cmyk}{0,.0675,.15,0}

\definecolor{KITred}{rgb}{.63,.13,.13}
\definecolor{kit-red100}{cmyk}{.25,1,1,0}
\definecolor{kit-red70}{cmyk}{.175,.7,.7,0}
\definecolor{kit-red50}{cmyk}{.125,.5,.5,0}
\definecolor{kit-red30}{cmyk}{.075,.3,.3,0}
\definecolor{kit-red15}{cmyk}{.0375,.15,.15,0}

\definecolor{KITpurple}{RGB}{160,0,120}
\colorlet{kit-purple100}{KITpurple}
\colorlet{kit-purple70}{KITpurple!70}
\colorlet{kit-purple50}{KITpurple!50}
\colorlet{kit-purple30}{KITpurple!30}
\colorlet{kit-purple15}{KITpurple!15}

\definecolor{KITcyanblue}{RGB}{80,170,230}
\colorlet{kit-cyanblue100}{KITcyanblue}
\colorlet{kit-cyanblue70}{KITcyanblue!70}
\colorlet{kit-cyanblue50}{KITcyanblue!50}
\colorlet{kit-cyanblue30}{KITcyanblue!30}
\colorlet{kit-cyanblue15}{KITcyanblue!15}

\usepackage[font=footnotesize]{subcaption}          %
\usepackage{booktabs}
\usepackage{dsfont}
\usepackage{stackrel}

\usepackage[nonumberlist, acronym]{glossaries}
\makeglossaries
\newacronym{SNN}{SNN}{spiking neural network}
\newacronym{LIF}{LIF}{leaky integrate-and-fire}
\newacronym{LI}{LI}{leaky integrate}
\newacronym{LDPC}{LDPC}{Low-density parity-check}
\newacronym{BP}{BP}{belief propagation}
\newacronym{DD-BMP}{DD-BMP}{differential decoding with binary message passing}
\newacronym{SPA}{SPA}{sum-product algorithm}
\newacronym{SR}{SR}{successive relaxation}
\newacronym{MS}{MS}{min-sum}
\newacronym{NMS}{NMS}{normalized MS}
\newacronym{ELENA}{ELENA-SNN}{enlarge-likelihood-each-notable-amplitude spiking-neural-network}
\newacronym{FG}{FG}{finite-geometry}
\newacronym{MLE}{ML-ELENA-SNN}{multi-level ELENA-SNN}

\begin{document}

\title{Spiking Neural Belief Propagation Decoder for LDPC Codes with Small Variable Node Degrees
\thanks{
This work has received funding from the European Research Council (ERC) under the European Union's Horizon 2020 research and innovation program (grant agreement No. 101001899). Parts of this work were carried out in the framework of the CELTIC-NEXT project AI-NET-ANTILLAS (C2019/3-3) (grant agreement 16KIS1316) and within the project Open6GHub (grant agreement 16KISK010) funded by the German Federal Ministry of Education and Research (BMBF).}
}

\author{\IEEEauthorblockN{Alexander von Bank, Eike-Manuel Edelmann, Jonathan Mandelbaum, and Laurent Schmalen}
\IEEEauthorblockA{Communications Engineering Lab, Karlsruhe Institute of Technology, 76187 Karlsruhe, Germany \\
(email: \texttt{\{alexander.bank, edelmann, jonathan.mandelbaum, laurent.schmalen\}@kit.edu})}
}

\maketitle

\begin{abstract}
Spiking neural networks (SNNs) promise energy-efficient data processing by imitating the event-based behavior of biological neurons.
In previous work, we introduced the enlarge-likelihood-each-notable-amplitude spiking-neural-network (ELENA-SNN) decoder, a novel decoding algorithm for low-density parity-check (LDPC) codes. 
The decoder integrates SNNs into belief propagation (BP) decoding by approximating the check node (CN) update equation using SNNs.
However, when decoding LDPC codes with a small variable node (VN) degree, the approximation gets too rough, and the ELENA-SNN decoder does not yield good results.
This paper introduces the multi-level ELENA-SNN (ML-ELENA-SNN) decoder, which is an extension of the ELENA-SNN decoder.
Instead of a single SNN approximating the CN update, multiple SNNs are applied in parallel, resulting in a higher resolution and higher dynamic range of the exchanged messages.
We show that the ML-ELENA-SNN decoder performs similarly to the ubiquitous normalized min-sum decoder for the $(38400,\,30720)$ regular LDPC code with a VN degree of $d_v=3$ and a CN degree of $d_\mathrm{c}=15$. 
\end{abstract}


\vspace{-.2cm}
\section{Introduction}
\vspace{-.1cm}
\gls{LDPC} codes are forward error-correcting codes that have been integrated into various modern communication standards, such as 5G and WLAN, due to their excellent error-correction capabilities when using iterative message-passing decoding, often known as \gls{BP} decoding. 
In particular, the \gls{SPA} together with \gls{LDPC} codes can achieve performance close to the channel capacity~\cite{RS01}. 
However, the \gls{SPA} involves computationally demanding check node (CN) updates~\cite{KLF2001}. 
To reduce this complexity, simplified methods like the \gls{MS} algorithm~\cite{FMI99}, the \gls{NMS} algorithm~\cite{AVAT2014}, and \gls{DD-BMP}~\cite{MBH2009} have been proposed.
In~\cite{CDEFX2005}, it was shown that suppressing small values during \gls{MS} decoding can boost performance.
By integrating memory into the decoding procedure, the decoding performance of the \gls{SPA} and \gls{MS} decoder can be further enhanced~\cite {XTB2008}.

One of the most powerful and efficient processing units is the human brain, solving challenging real-world problems while only consuming roughly \SI{25}{\watt} of power.
Neuromorphic engineering tries to build scalable, low-power systems by implementing the essential features of the brain, allowing an efficient emulation of brain-inspired neural networks~\cite{ZN21}. 
Most recent neuromorphic hardware emulates~\glspl{SNN}~\cite{Intel2021, PBXKSSWLMS22}, which mimic the event-based behavior of biological neurons~\cite[pp.~12,13]{GK02}.

In~\cite{ELENA}, we introduced the \gls{ELENA} decoder, which approximates computational heavy parts of the \gls{SPA} decoder by \glspl{SNN}.
While a threshold operation suppresses messages with small amplitudes in an all-or-nothing manner, spiking neurons introduce memory to the decoder; both can enhance the decoding performance.
In~\cite{ELENA}, two binary $(N, k)$ LDPC codes, which encode $k$ data bits into $N$ code bits, were used: the $(273,\,191)$ and $(1023,\,781)$ \gls{FG} \gls{LDPC} codes. 
For both codes, we showed that the \gls{ELENA} decoder achieves approximately the performance of the \gls{NMS} decoder with \gls{SR}, outperforming other complexity-reduced versions of the \gls{SPA} decoder and even outperforming the \gls{SPA} decoder in high SNR regimes~\cite{ELENA}. 
To the best of our knowledge, the \gls{ELENA} decoder is the first \gls{SNN}-based decoder. 

However, we observed that the \gls{ELENA} decoder yields poor results when decoding \gls{LDPC} codes with low variable node (VN) degrees. 
The combination of the threshold operation and the low VN degrees leads to a low dynamic range and low resolution of the messages. 

This paper expands the \gls{ELENA} decoder to the \gls{MLE} decoder.
Multiple thresholds are used instead of a single threshold, introducing a finer granularity of the message levels and a wider dynamic range of the messages.
For an \gls{LDPC} code with low VN degree $d_\mathrm{v}=3$, we show that the \gls{MLE} decoder performs close to the \gls{NMS} decoder, while the \gls{ELENA} decoder does not show competitive performance.
The code for reproducing the results is available at~\cite{github-code}.

\section{Spiking Neural Networks}
\begin{figure}
    \centering
    \resizebox{0.49\textwidth}{!}{
    \input{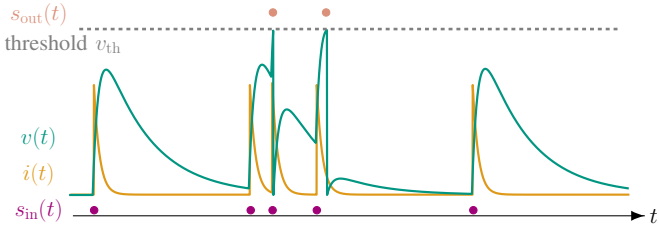}
    }
    \caption{Example of the dynamics of an \gls{LIF} neuron. The input spikes $s_\mathrm{in}(t)$, denoted by purple dots, induce the synaptic current $i(t)$, which charges the membrane potential $v(t)$. If $v(t)$ exceeds the threshold $v_\mathrm{th}$, an output spike $s_\mathrm{out}(t)$ is fired, and $v(t)$ is reset.}
    \label{fig:lif_dynamics}
\end{figure}

\glspl{SNN} are networks of interconnected, \emph{spiking neurons} with a state which try to mimic the event-based signal processing of the mammalian brain. 
The exchange of information between the neurons consists of short, uniform pulses, so-called spikes.
A sequence of spikes $s(t)\in \{0,1\}$ is called a spike train, where $t\in\mathbb{R}$ is the time.
Like the biological neuron, a spiking neuron receives inputs, i.e., spike trains, from upstream connected neurons, processes those, and generates an output spike train $s_\mathrm{out}(t)$~\cite[p.~13]{GK02}.
The connection between two neurons is called a synapse and possesses a weight, which defines the strength of the connection~\cite[p.~14]{GK02}. 
Hence, each incoming spike train $s_j(t)$ is scaled by a weight ${w_j\in\mathbb{R}}$, where $s_j(t)$ denotes the spike train received by the $j$-th upstream connected neuron and $w_j$ models the connecting synapse.  
The incoming spike trains induce the \emph{synaptic current} $i(t)$, which then charges the internal state of the neuron, the \emph{membrane potential} $v(t)$. 
In parallel, the neuron leaks towards its initial state $v_\mathrm{rest}$.
Hence, the neuron can be described as a leaky integrator of its input signals~\cite{NMZ2019}.
If $v(t)$ exceeds a fixed threshold $v_\mathrm{th}$, the neuron emits an output spike, and the membrane potential $v(t)$ is reset to $v_\mathrm{rest}$.
A widely used spiking neuron model is the \gls{LIF} neuron model since it captures the fundamental concepts of the biological neuron while being computationally attractive.
At discrete time instants $t=\kappa \Delta t,\, \kappa \in \mathbb{N}$, the discrete dynamics of the \gls{LIF} neuron are~\cite{NMZ2019} 
\begin{align*}
    v[\kappa+1] &= v[\kappa] \cdot \mathrm{e}^{-\frac{\Delta t}{\tau_\mathrm{m}}} + i[\kappa] \cdot \mathrm{e}^{-\frac{\Delta t}{\tau_\mathrm{m}}} \, ,\\[0.5em]
    i[\kappa +1] &= i[\kappa] \cdot \mathrm{e}^{-\frac{\Delta t}{\tau_\mathrm{s}}} + \sum_j s_j[\kappa] w_j \, , 
\end{align*}
where $\Delta t$ is the sampling time of the system, $\tau_\mathrm{m}$ denotes the time constant describing the leakage of the membrane potential, and $\tau_\mathrm{s}$ describes the temporal dynamics of the processes of the synapse.  
Output spikes are generated according to
\begin{align*}
    s_\mathrm{out}[\kappa] = \Theta\left(v[\kappa]-v_\mathrm{th} \right) = \begin{cases}
        1, &\text{if}\quad v[\kappa] > v_\mathrm{th}\, , \\
        0, &\text{otherwise} \, ,
    \end{cases}
\end{align*}
where $\Theta(\cdot)$ denotes the Heaviside step function.
If an output spike is generated, the state of the neuron is reset, i.e., ${v[\kappa] \leftarrow v_\mathrm{rest}}$.
Fig.~\ref{fig:lif_dynamics} shows an exemplary realization of the \gls{LIF} neuron dynamics.
Another common spiking neuron model is the \gls{LI} neuron model.
It exhibits the same temporal dynamics as the \gls{LIF} neuron;
however, the spiking functionality is deactivated. 
Hence, it does not generate an output spike, and the \gls{LI} neuron acts as a temporary memory.
In this paper, we use the PyTorch-based framework \textit{Norse}~\cite{Norse} for the simulation of \glspl{SNN}. 

\section{Belief Propagation Decoding}
\label{sec:Belief Propagation Decoder}
We focus on the decoding of binary $(N, k)$ LDPC codes. These LDPC codes are defined by the null space of a sparse parity-check matrix (PCM) $\bm{H} \in \mathbb{F}_2^{M \times N}$. 
The SPA operates on the Tanner graph, which is a bipartite graph associated with the PCM~\cite[pp.~51-59]{RU2008}.
A Tanner graph contains two types of nodes: CNs $\mathsf{c}_j,\, {j \in \{1, 2, \ldots, M\}}$, each representing a single parity check or a row of the PCM, and VNs $\mathsf{v}_i, \,i \in \{1, 2, \ldots, N\}$, each associated with a code bit, corresponding to a column of the PCM. 
An edge between CN $\mathsf{c}_j$ and VN $\mathsf{v}_i$ exists if and only if $H_{j,i} = 1$. A regular ($d_\mathrm{v}$, $d_\mathrm{c}$) LDPC code is characterized by having $d_\mathrm{c}$ non-zero entries in each row of the PCM and $d_\mathrm{v}$ non-zero entries in each column of its PCM, where $d_\mathrm{c}$ and $d_\mathrm{v}$ denote the CN and VN degrees, respectively.

Consider a codeword $\bm{x}=(x_1,\ldots,x_N)$ modulated using binary phase shift keying (BPSK) and transmitted over a binary-input additive white Gaussian noise (AWGN) channel, described as ${y_i = (-1)^{x_i} + n_i}$, with $n_i \sim \mathcal{N}(0, \sigma^2)$.

The bit-wise log-likelihood ratio (LLR) $L_i$ at the channel output, related to the $i$-th code bit, is defined as 
\begin{align}
    L_i = \log{\left(\frac{P(Y_i=y_i|X_i=0)}{P(Y_i=y_i|X_i=1)}\right)}.
    \label{eq:likelihood}
\end{align}

Next, we employ the SPA, in which messages are represented as LLRs, iteratively updated in the nodes, and propagated along the edges of the Tanner graph. An SPA iteration using a flooding schedule involves the parallel update of all CN messages, followed by updating all VN messages. 
The messages from VNs to CNs are initialized as ${L_{i \rightarrow j}^{[\mathrm{v}]} = L_i,\, \forall j \in\mathcal{N}(i)}$, where $L_{i \rightarrow j}^{[\mathrm{v}]}$ denotes the message sent from VN $\mathsf{v}_i$ to CN $\mathsf{c}_j$ and $\mathcal{N}(i) = \{j : H_{j,i} = 1\}$ is the set of indices of CNs connected to VN $\mathsf{v}_i$.

First, the check-to-variable-node messages $L_{i \leftarrow j}^{[\mathrm{c}]}$ are computed. 
The update equation~{\cite[Eq.~(2.17)]{RU2008}} can be simplified, by splitting $L_{i\leftarrow j}^{[\mathrm{c}]}$ into its absolute value~${\alpha_{i \leftarrow j}^{[\mathrm{c}]} = \vert L_{i \leftarrow j}^{[\mathrm{c}]} \vert}$ and its sign~${\beta_{i \leftarrow j}^{[\mathrm{c}]} = \mathrm{sign} ( L_{i \leftarrow j}^{[\mathrm{c}]} )}$, such that~${{L_{i\leftarrow j}^{[\mathrm{c}]} = \alpha_{i\leftarrow j}^{[\mathrm{c}]} ~ \beta_{i\leftarrow j}^{[\mathrm{c}]} ,\,\forall i \in \mathcal{M}(j)}}$, with
\begin{align}
        \alpha_{i\leftarrow j}^{[\mathrm{c}]} &=  2\cdot \textrm{tanh}^{-1}
                                    \left(
                                        \prod_{i' \in \mathcal{M}(j)\setminus \{i\}}
                                        \textrm{tanh}\left(\frac{\big|L_{i'\rightarrow j}^{[\mathrm{v}]}\big|}{2}\right) 
                                    \right)\label{eq:cn_update_alpha},\\[.1em]
    \beta_{i\leftarrow j}^{[\mathrm{c}]} &= \prod_{i' \in \mathcal{M}(j)\setminus \{i\}} 
                                    \textrm{sign}\left( L_{i'\rightarrow j}^{[\mathrm{v}]} \right) \label{eq:cn_update_beta}  ,
\end{align}
where $\mathcal{M}(j) = \{i : H_{j,i} = 1\}$ is the set of indices of VNs connected to CN $j$.

Next, the VN update involves calculating the message from $\mathsf{v}_i$ to $\mathsf{c}_j$ using 
\begin{align}
    L_{i\rightarrow j}^{[\mathrm{v}]} &= L_i + 
    \hspace*{-3mm}
    \sum_{j'\in \mathcal{N}(i)\setminus \{j\}} L_{i\leftarrow j'}^{[\mathrm{c}]} ,\quad \forall j \in \mathcal{N}(i)\,
    \label{eq:vn_update}.
\end{align}

Both the CN and VN updates adhere to the extrinsic principle, meaning that when updating the message to VN $\mathsf{v}_i$ (or CN $\mathsf{c}_j$), the message originating from VN~$\mathsf{v}_i$ (or CN $\mathsf{c}_j$) is excluded from the update equation, hence, $i' \in \mathcal{M}(j)\setminus \{i\}$. After reaching a predefined maximum number of iterations, the bit-wise output LLRs $\tilde{L}_i$ are computed as 
\begin{align}
    \tilde{L}_{i} &= L_i + \sum_{j\in \mathcal{N}(i)} L_{i\leftarrow j}^{[\mathrm{c}]} \; .
    \label{eq:Estimate}
\end{align}
\vspace*{-3mm}\\
Finally, the decoded bits $\hat{b}_i$ are obtained by applying a hard decision $\hat{b}_i = \mathbbm{1}_{\{\tilde{L}_i \leq 0\}}$.
For a more detailed discussion of message-passing algorithms, we refer the reader to~\cite[Ch.~2]{RU2008}.

\section{\gls{ELENA} Decoder} 
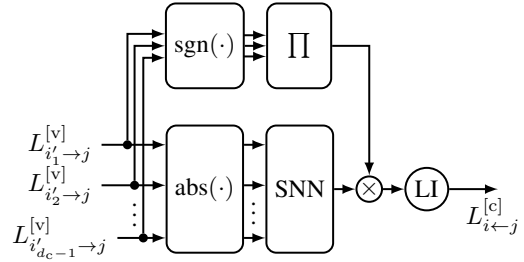
\begin{figure}[t!] 
    \begin{center} 
    \resizebox{0.4\textwidth}{!}{ \begin{tikzpicture}[>=latex,thick]
    \def\ydist{0.5cm}
    \def\xdist{0.3cm}
    \def\h{1.2cm}
    \node[] (L2) {$L_{i_2'\rightarrow j}^{[\mathrm{v}]}$};
    \node[above=-0.25cm of L2] (L1) {$L_{i_1'\rightarrow j}^{[\mathrm{v}]}$};
    \node[below=-0.1cm of L2] (L3) {$L_{i_{d_\mathrm{c}-1}'\rightarrow j}^{[\mathrm{v}]}$};

    \node[above right=-1.38cm and 3*\xdist of L2,draw,rectangle,rounded corners, minimum height=1.5*\h] (abs) {abs$(\cdot)$};
    \node[right=\xdist of abs,draw,rectangle, rounded corners,minimum height=1.5*\h,minimum width=0.9cm] (snn) {SNN};
    \node[right=\xdist of snn,draw,circle, inner sep=0pt,radius=0.1cm] (m1) {$\times$};

    \draw[->] (L1) to (\tikztostart -| abs.west);
    \draw[->] (L2) to (\tikztostart -| abs.west);
    \draw[->] (L3) to (\tikztostart -| abs.west);

    \draw[->] (L1 -| abs.east) to (\tikztostart -| snn.west);
    \draw[->] (L2 -| abs.east) to (\tikztostart -| snn.west);
    \draw[->] (L3 -| abs.east) to (\tikztostart -| snn.west);

    \draw[->] (snn) -- (m1);

    \node[right=\xdist of L1,draw,circle,inner sep=1, fill=black] (b1) {};
    \node[right=\xdist+0.1cm of L2,draw,circle,inner sep=1, fill=black] (b2) {};
    \node[right=\xdist       of L3,draw,circle,inner sep=1, fill=black] (b3) {};
    \node[below=-0.15cm of b2] () {$\vdots$};
    \node[below right =-0.15cm and 1.45cm of b2] () {$\vdots$};

    \node[above=\ydist of abs,draw, rectangle, rounded corners,minimum height=\h] (sgn) {sgn$(\cdot)$};
    \node[right=\xdist of sgn,draw,rectangle,rounded corners,,minimum height=\h,minimum width=0.9cm] (product) {$\prod$};

    \draw[->] (b1) -- +(0cm,1.57cm) to (\tikztostart -|sgn.west);
    \draw[->] (b2) |- (sgn);
    \draw[->] (b3) -- +(0cm,2.55cm) to (\tikztostart -|sgn.west);

    \draw[->] (sgn)++(0.55cm,0.15cm) to (\tikztostart -| product.west);
    \draw[->] (sgn) to (\tikztostart -| product.west);
    \draw[->] (sgn)++(0.55cm,-0.15cm) to (\tikztostart -| product.west);

    \draw[->] (product) -| (m1);

    \node[right=\xdist of m1,draw,circle, inner sep=2pt] (LI) {LI};
    \draw[->] (m1) -- (LI);

    \draw[->] (LI) -- ++(1cm,0cm) node[pos=0.9,below] () {$L_{i \leftarrow j}^{[\mathrm{c}]}$};

\end{tikzpicture} } \caption{\gls{ELENA} SCNU that computes the message $L_{i\leftarrow j}^{[\mathrm{c}]}$ based on the incoming messages~${L^{[\mathrm{v}]}_{i'\rightarrow j},~i'\in \mathcal{M}(j)\setminus\{i\}=:\{i_1,\ldots,i_{d_\mathrm{c}-1}\}}$. The top and bottom branches approximate~\eqref{eq:cn_update_alpha} and~\eqref{eq:cn_update_beta}, respectively. The LI neuron serves as the memory element.} 
    \label{fig:SPC_SRNCTMS} 
    \end{center} 
\end{figure}

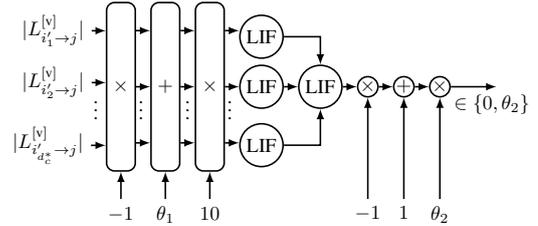
\begin{figure}[t!] 
    \begin{center} 
    \resizebox{0.4\textwidth}{!}{ \begin{tikzpicture}[>=latex,thick]
    \def\ydist{0.5cm}
    \def\xdist{0.25cm}
    \def\h{1.2cm}
    \node[] (L2) {$|L_{i_2'\rightarrow j}^{[\textrm{v}]}|$};
    \node[above= 0.1cm of L2] (L1) {$|L_{i_1'\rightarrow j}^{[\textrm{v}]}|$};
    \node[below left= 0.25cm and -1.35cm of L2] (L3) {$|L_{i_{d^\ast_c}'\rightarrow j}^{[\textrm{v}]}|$};

    \node[above right=-2.0cm and 1*\xdist of L2,draw,rectangle,rounded corners, minimum height=2.5*\h] (mul1) {$\times$};
    \draw[->] (mul1)++(0cm,-2.0cm) -- (mul1.south);
    \node[below=0.5cm of mul1] () {$-1$};

    \node[right=\xdist of mul1,draw,rectangle,rounded corners, minimum height=2.5*\h] (add1) {$+$};
    \draw[->] (add1)++(0cm,-2.0cm) -- (add1.south);
    \node[below=0.48cm of add1] () {$\theta_1$};

    \node[right=\xdist of add1,draw,rectangle,rounded corners, minimum height=2.5*\h] (mul2) {$\times$};
    \draw[->] (mul2)++(0cm,-2.0cm) -- (mul2.south);
    \node[below=0.5cm of mul2] () {$10$};
    
    \node[right=\xdist of mul2 ,draw,circle, inner sep=2pt,radius=2.0cm] (LIF2) {LIF};
    \node[above= 0.1cm of LIF2 ,draw,circle, inner sep=2pt,radius=2.0cm] (LIF1) {LIF};
    \node[below= 0.25cm of LIF2,draw,circle, inner sep=2pt,radius=2.0cm] (LIF3) {LIF};

    \node[right=\xdist of LIF2 ,draw,circle, inner sep=2pt,radius=2.0cm] (LIFOR) {LIF};
    
    \node[right=\xdist of LIFOR,draw,circle, inner sep=0pt,radius=2.0cm] (mul3) {$\times$};
    \draw[->] (mul3)++(0cm,-2.0cm) -- (mul3.south);
    \node[below=1.8cm of mul3] () {$-1$};
    
    \node[right=\xdist of mul3,draw,circle, inner sep=0pt,radius=2.0cm] (add3) {$+$};
    \draw[->] (add3)++(0cm,-2.0cm) -- (add3.south);
    \node[below=1.8cm of add3] () {$1$};
    
    \node[right=\xdist of add3,draw,circle, inner sep=0pt,radius=2.0cm] (mul4) {$\times$};
    \draw[->] (mul4)++(0cm,-2.0cm) -- (mul4.south);
    \node[below=1.8cm of mul4] () {$\theta_2$};

    \draw[->] (L1) to (\tikztostart -| mul1.west);
    \draw[->] (L2) to (\tikztostart -| mul1.west);
    \draw[->] (L3) to (\tikztostart -| mul1.west);

    \draw[->] (mul1.east) to (add1.west);
    \draw[->] (mul1.east)++(0,1cm) -- ++(\xdist,0cm);
    \draw[->] (mul1.east)++(0,-1cm) -- ++(\xdist,0cm);

    \draw[->] (add1.east) to (mul2.west);
    \draw[->] (add1.east)++(0,1cm) -- ++(\xdist,0cm);
    \draw[->] (add1.east)++(0,-1cm) -- ++(\xdist,0cm);

    \draw[->] (mul2.east) to (LIF2.west);
    \draw[->] (mul2.east)++(0,1cm) -- ++(\xdist,0cm);
    \draw[->] (mul2.east)++(0,-1cm) -- ++(\xdist,0cm);

    \draw[->] (LIF1.east) -| (LIFOR.north);
    \draw[->] (LIF2.east) -- (LIFOR.west);
    \draw[->] (LIF3.east) -| (LIFOR.south);

    \draw[->] (LIFOR) -- (mul3);
    \draw[->] (mul3) -- (add3);
    \draw[->] (add3) -- (mul4);

    \draw[->] (mul4) -- ++(1.0cm,0cm) node[pos=0.9,below] () {$\in\{0,\theta_2\}$};

    \node[above right=-1.2cm and -0.1cm of L2] () {$\vdots$};
    \node[above right=-1.2cm and 0.7cm of L2] () {$\vdots$};
    \node[above right=-1.2cm and 1.52cm of L2] () {$\vdots$};
    \node[above right=-1.2cm and 2.27cm of L2] () {$\vdots$};
    
\end{tikzpicture} } \caption{Structure of the SNN block from Fig.~\ref{fig:SPC_SRNCTMS}, implementing~\eqref{eq:elena_offset} using $d_\mathrm{c}$ \gls{LIF} neurons. If an input falls below $\theta_1$, the connected neuron is charged to trigger a spike. The upstream neuron aggregates all incoming spikes and forwards one, resulting in a zero output. If all inputs exceed $\theta_1$, no neurons are charged, producing $\theta_2$ as the output.} 
    \label{fig:SNN} 
    \end{center} 
\end{figure}

A significant part of the computational complexity of the SPA lies in the CN update~\eqref{eq:cn_update_alpha}~\cite{KLF2001}. 
In~\cite{ELENA}, we introduced the \gls{ELENA} decoder, which approximates the CN update using SNNs. 
\gls{ELENA} employs an SNN-based CN update (SCNU) to incorporate SNNs into the CN update~\eqref{eq:cn_update_alpha}-\eqref{eq:cn_update_beta}. 
Fig.~\ref{fig:SPC_SRNCTMS} illustrates the setup of an SCNU that computes the message $L^{[\mathrm{c}]}_{i\leftarrow j}$ based on the messages $L^{[\mathrm{v}]}_{i'\rightarrow j},\,i'\in \mathcal{M}(j)\setminus\{i\}$. 
The incoming messages $L_{i'\rightarrow j}^{[\mathrm{v}]}$ are decomposed into their sign and absolute value, which are then processed separately, combined, and integrated over time by an LI neuron, which acts as memory. 
Thus, the parameters $\tau_\mathrm{m}$ and $\tau_\mathrm{s}$ of the LI neuron control the memory behavior of the SCNU. 
The top branch performs the operation in~(\ref{eq:cn_update_beta}), which can be executed using XOR operations, while the bottom branch implements~(\ref{eq:cn_update_alpha}) with the SNN shown in Fig.~\ref{fig:SNN}.

The offset \gls{MS} algorithm~\cite{CDEFX2005}\footnote{There exist two perspectives on the \gls{MS} algorithm. 
On the one hand, it can be viewed as an approximation of SPA, which is the belief propagation algorithm (BPA) on the sum-product semiring. On the other hand, the MS algorithm can be derived as the BPA on a different semiring, namely the max-product semiring, yielding the message-passing algorithm that maximizes the block-wise maximum a-posteriori probability (MAP) \cite[p. 63]{RU2008}. Due to cycles, both algorithms are suboptimal. Even though SPA typically yields better decoding performance depending on the graph structure, refined versions of the MS algorithm, e.g., NMS, can outperform SPA.}
, which replaces~\eqref{eq:cn_update_alpha} by ${{\alpha_{i\leftarrow j}^{[\mathrm{c}]}\approx \max \left\{\mathrm{min}_{i' \in \mathcal{M}(j)\setminus \{i\}}\, \big\{\big\vert L_{i'\rightarrow j}^{[\mathrm{v}]} \big\vert - \theta_1\big\} ,0  \right\}}}$, inspired us to further simplify the computation via
\begin{align} 
    \alpha_{i\leftarrow j}^{[\mathrm{c}]}\approx \begin{cases}
        \theta_2, &\text{if}\; \min_{i' \in \mathcal{M}(j)\setminus \{i\}} \, \left\vert L_{i'\rightarrow j}^{[\mathrm{v}]} \right\vert > \theta_1\, , \\
        0, &\text{otherwise}\, ,
    \end{cases} \
\label{eq:elena_offset}
\end{align}
with $\theta_1$ being a threshold.
Similar to the offset \gls{MS} algorithm,~(\ref{eq:elena_offset}) outputs zero if the input values are below the threshold $\theta_1\in\mathbb{R}^{+}$. 
Unlike the offset \gls{MS} algorithm, which outputs the biased minimum,~(\ref{eq:elena_offset}) returns a fixed value ${\theta_2\in\mathbb{R}^{+}}$ once the threshold $\theta_1$ is exceeded. 
Fig.~\ref{fig:SNN} illustrates how \gls{LIF} neurons implement~(\ref{eq:elena_offset}). 
Initially, the signs of all inputs ${\vert L_{i'\rightarrow j}^{[\mathrm{v}]} \vert}$ are reversed, and after adding the bias $\theta_1$, the signal is amplified by $10$ and integrated by \gls{LIF} neurons. 
If ${(\theta_1~-~\vert L_{i'\rightarrow j}^{[\mathrm{v}]}\vert) >0}$, the corresponding \gls{LIF} neuron is charged, causing an output spike that is passed to the combining \gls{LIF} neuron. 
If at least one spike reaches the combining neuron, it generates a spike, yielding an overall output of zero. If $(\theta_1 - \vert L_{i'\rightarrow j}^{[\mathrm{v}]} \vert) <0$, the corresponding \gls{LIF} neuron remains inactive, and no output spike is generated. Consequently, if no \gls{LIF} neuron fires, the combining neuron remains silent, resulting in $\theta_2$ as output.

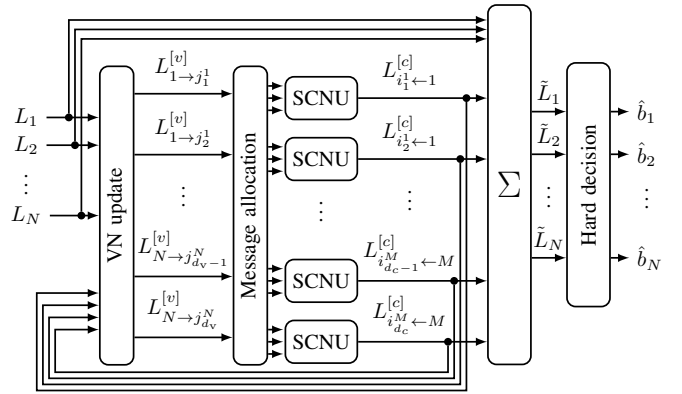
\begin{figure}[t!] 
    \begin{center} 
    \resizebox{0.5\textwidth}{!}{ \begin{tikzpicture}[>=latex, thick]
\def\ydist{0.5cm}
        \def\xdist{.8cm}
    
        \node [] (L1) {$L_1$};
        \node[below = -0.1cm of L1] (L2) {$L_2$};
        \node [below of =  L1] (h) {$\vdots$};
        \node[below = 1.2*\ydist of L2] (LN) {$L_N$};
    
        \node[right=1*\xdist of LN, draw, rectangle, rounded corners, align=center,minimum height = 4.9cm] (VN_up) {\rotatebox{90}{VN update}};  
        \draw[->] (L1) to (\tikztostart -| VN_up.west);
        \draw[->] (L2) to (\tikztostart -| VN_up.west);
        \draw[->] (LN) to (\tikztostart -| VN_up.west);
    
        \node[right=2*\xdist of VN_up, draw, rectangle, rounded corners, minimum height = 4.9cm] (dist) {\rotatebox{90}{Message allocation}};
        \draw[->] (VN_up.east)++(0cm,2cm) -- +(2*\xdist,0cm) node[midway, above] {$L_{1 \rightarrow j_1^1}^{[v]}$ };        
        \draw[->] (VN_up.east)++(0cm,1cm) -- +(2*\xdist,0cm) node[midway, above] (h1) {$L_{1 \rightarrow j_2^1}^{[v]}$};
        \draw[->] (VN_up.east)++(0cm,-1cm) -- +(2*\xdist,0cm) node[midway, above] ()
        {$L_{N \rightarrow j_{d_\mathrm{v}-1}^N}^{[v]}$};
        \draw[->] (VN_up.east)++(0cm,-2cm) -- +(2*\xdist,0cm) node[midway, above] ()
        {$L_{N \rightarrow j_{d_\mathrm{v}}^N}^{[v]}$};
        \node[below of=h1] () {$\vdots$};
        
        \def\h{0.7cm}
        \node [above right = -0.9cm and \xdist/3 of dist, draw, rectangle, rounded corners, minimum height = \h] (SCNU1) {SCNU};
        \node [above right = -1.9cm and \xdist/3 of dist, draw, rectangle, rounded corners, minimum height = \h] (SCNU2) {SCNU};
        \node [above right = -3.9cm and \xdist/3 of dist, draw, rectangle, rounded corners, minimum height = \h] (SCNU3) {SCNU};
        \node [above right = -4.9cm and \xdist/3 of dist, draw, rectangle, rounded corners, minimum height = \h] (SCNUM) {SCNU};
        \node [below=of SCNU1] (h2) {$\vdots$};

        \def\yoff{0.2cm}
        \draw[<-] (SCNU1)+(-0.62cm,\yoff) to (\tikztostart -| dist.east);
        \draw[<-] (SCNU1)+(-0.62cm,0.0cm) to (\tikztostart -| dist.east);
        \draw[<-] (SCNU1)+(-0.62cm,-\yoff) to (\tikztostart -| dist.east);

        \draw[<-] (SCNU2)+(-0.62cm,\yoff) to (\tikztostart -| dist.east);
        \draw[<-] (SCNU2)+(-0.62cm,0.0cm) to (\tikztostart -| dist.east);
        \draw[<-] (SCNU2)+(-0.62cm,-\yoff) to (\tikztostart -| dist.east);

        \draw[<-] (SCNU3)+(-0.62cm,\yoff) to (\tikztostart -| dist.east);
        \draw[<-] (SCNU3)+(-0.62cm,0.0cm) to (\tikztostart -| dist.east);
        \draw[<-] (SCNU3)+(-0.62cm,-\yoff) to (\tikztostart -| dist.east);
        
        \draw[<-] (SCNUM)+(-0.62cm,\yoff) to (\tikztostart -| dist.east);
        \draw[<-] (SCNUM)+(-0.62cm,0.0cm) to (\tikztostart -| dist.east);
        \draw[<-] (SCNUM)+(-0.62cm,-\yoff) to (\tikztostart -| dist.east);

        \def\x{2.1cm}
        \node [above right=-1.75cm and \x of SCNU3, draw, rectangle, rounded corners, minimum height=5.9cm,minimum width =0.7cm] (sum) {\LARGE $\Sigma$};
        \draw[->] (SCNU1) -- +(\x+0.6cm,0cm) node[pos=0.4, above] () 
        {$L_{i^1_1 \leftarrow 1}^{[c]}$}; 
        \draw[->] (SCNU2) -- +(\x+0.6cm,0cm) node[pos=0.4, above] (hc) {$L_{i^1_2 \leftarrow 1}^{[c]}$};
        \draw[->] (SCNU3) -- +(\x+0.6cm,0cm) node[pos=0.4, above] () {$L_{i^M_{d_c-1} \leftarrow M}^{[c]}$};
        \draw[->] (SCNUM) -- +(\x+0.6cm,0cm) node[pos=0.4, above] () {$L_{i^M_{d_c} \leftarrow M}^{[c]}$};
        \node[below of=hc] () {$\vdots$};

        \node [right=0.3cm of L1, draw, circle, inner sep =1pt, fill=black] (n1) {};
        \node [right = 0.4 of L2, draw, circle, inner sep =1pt, fill=black] (n2) {};
        \node [right=0.45cm of LN, draw, circle, inner sep =1pt, fill=black] (n3) {};
        \draw[->] (n1) -- +(0cm,1.6cm) to (\tikztostart -| sum.west);
        \draw[->] (n2) -- +(0cm,1.9cm) to (\tikztostart -| sum.west);
        \draw[->] (n3) -- +(0cm,2.9cm) to (\tikztostart -| sum.west);

        \node[right=\x-0.4cm of SCNU1, draw, fill = black, circle, inner sep =1pt] (m1) {};
        \node[right=\x-0.5cm of SCNU2, draw, fill = black, circle, inner sep =1pt] (m2) {};
        \node[right=\x-0.6cm of SCNU3,draw, fill = black,circle, inner sep=1pt] (m3) {};
        \node[right=\x-0.7cm of SCNUM, draw, fill = black, circle, inner sep =1pt] (m4) {};
        \draw[->] (m1) -- +(0cm,-4.8cm) -- +(-7.0cm,-4.8cm) -- +(-7.0cm,-3.2cm) to (\tikztostart -| VN_up.west);
        \draw[->] (m2) -- +(0cm,-3.7cm) -- +(-6.8cm,-3.7cm) -- +(-6.8cm,-2.4cm) to (\tikztostart -| VN_up.west);
        \draw[->] (m3) -- +(0cm,-1.6cm) -- +(-6.6cm,-1.6cm) -- +(-6.6cm,-0.6cm) to (\tikztostart -| VN_up.west);
        \draw[->] (m4) -- +(0cm,-0.5cm) -- +(-6.4cm,-0.5cm) -- +(-6.4cm,0.2cm) to (\tikztostart -| VN_up.west);

        \node [right=\xdist*0.7 of sum, draw, rectangle, rounded corners, minimum height = 4cm, minimum width =0.7cm] (dec) {\rotatebox{90}{Hard decision}};
        \draw[->] (sum.east)++(0cm,1.2cm) -- +(\xdist*0.7,0cm) node [pos=0.5, above] (h3) {$\tilde{L}_1$};
        \draw[->] (sum.east)++(0cm,0.5cm) -- +(\xdist*0.7,0cm) node [pos=0.5, above] {$\tilde{L}_2$};
        \draw[->] (sum.east)++(0cm,-1.2cm) -- +(\xdist*0.7,0cm) node [pos=0.5, above] {$\tilde{L}_N$};
        \node[below = 0.9cm of h3] () {$\vdots$};

        \draw[->] (dec.east)++(0cm,1.2cm) -- +(0.3cm,0cm) node [right] (h4) {$\hat{b}_1$};
        \draw[->] (dec.east)++(0cm,0.5cm) -- +(0.3cm,0cm) node [right] {$\hat{b}_2$};
        \draw[->] (dec.east)++(0cm,-1.2cm) -- +(0.3cm,0cm) node [right] {$\hat{b}_N$};
        \node[below = 0.6cm of h4] () {$\vdots$};
\end{tikzpicture} } 
    \caption{Architecture of the proposed decoder. The SCNU contains the SNN. For ${i\in \{1,\ldots,N\}}$ and ${j\in \{1,\ldots,M\}}$, the notation ${\mathcal{N}(i):=\{j_1^i,\ldots, j^i_{d_v}\}}$ and $ \mathcal{M}(j):=\{i_1^j,\ldots, i^j_{d_v}\}$ is used.} 
    \label{fig:SNN_BP_structure} 
\end{center} 
\end{figure}

Fig.~\ref{fig:SNN_BP_structure} shows the overall decoder architecture incorporating the SCNUs. All SCNUs share the same parameter set.
Initially, all $L_{i\leftarrow j}^{[\mathrm{c}]}$ are set to zero, and the VNs are updated using~\eqref{eq:vn_update}. Then, the iterative decoding begins. Each CN update involves passing the variable-to-check-node messages $L_{i\rightarrow j}^{[\mathrm{v}]}$ to the corresponding SCNU. For regular LDPC codes, there are $M\cdot d_{\mathrm{c}}$ SCNUs, each responsible for performing an extrinsic CN update. During message allocation, all variable-to-check-node messages $L_{i'\rightarrow j}^{[\mathrm{v}]}$, with ${i'\in \mathcal{M}(j)\setminus\{i\}}$, are routed to the respective SCNU. 
Following this, the VNs are updated, and the process is repeated iteratively until the hard decision maps the output LLRs to binary values.
Afterward, the membrane potentials and synaptic currents of all neurons are reset to zero, effectively clearing the memory.
Note that the decoder operates using a flooding schedule, meaning that all node updates are executed in parallel.
Each SCNU within the \gls{ELENA} decoder consists of~$d_\mathrm{c}$ \gls{LIF} spiking neurons and one LI neuron.

\section{Multilevel-ELENA}
\label{sec:ML-ELENA}
Due to the ternary output of the SCNU of \gls{ELENA}, i.e., $L_{i\leftarrow j}^{[\mathrm{c}]} \in \{-\theta_2,0,\theta_2\}$, the VN update value can be described by ${L_{i\rightarrow j}^{[\mathrm{v}]} = L_i \pm n \cdot \theta_2}$ with ${n\in\{-(d_\mathrm{v}-1),-(d_\mathrm{v}-2),\ldots,(d_\mathrm{v}-2),(d_\mathrm{v}-1)\}}$.
Thus, the resolution of $L_{i\rightarrow j}^{[\mathrm{v}]}$ is limited by $\theta_2$, and its maximum value highly depends on the initial value $L_i$ and the VN degree $d_\mathrm{v}$.
Hence, for codes with small VN degree $d_\mathrm{v}$, e.g., a regular (3, 6) \gls{LDPC} code, both the range and the resolution of $L_{i\rightarrow j}^{[\mathrm{v}]}$ are limited, resulting in performance limitations of the \gls{ELENA} decoder.

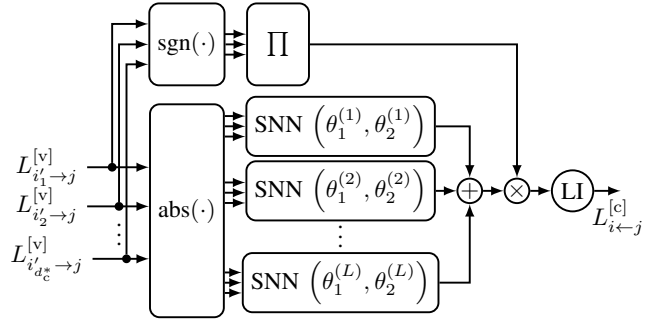
\begin{figure}
    \centering
    \resizebox{0.49\textwidth}{!}{
    \begin{tikzpicture}[>=latex,thick]
    \def\ydist{0.5cm}
    \def\xdist{0.3cm}
    \def\h{1.2cm}
    \node[] (L2) {$L_{i_2'\rightarrow j}^{[\mathrm{v}]}$};
    \node[above=-0.25cm of L2] (L1) {$L_{i_1'\rightarrow j}^{[\mathrm{v}]}$};
    \node[below=-0.1cm of L2] (L3) {$L_{i_{d^\ast_\mathrm{c}}'\rightarrow j}^{[\mathrm{v}]}$};

    \node[above right=-2.05cm and 3*\xdist of L2,draw,rectangle,rounded corners, minimum height=2.6*\h] (abs) {abs$(\cdot)$};
    \node[above right=-1.73cm and \xdist of abs,draw,rectangle, rounded corners,minimum height=.5*\h,minimum width=0.9cm] (snn) {SNN $\left({\theta_{1}^{(2)},\theta_{2}^{(2)}}\right)$};
    \node[above=.1*\ydist of snn,draw,rectangle, rounded corners,minimum height=.5*\h,minimum width=0.9cm] (snn0) {SNN $\left(\theta_{1}^{(1)},\theta_{2}^{(1)}\right)$};
    \node[below=.9*\ydist of snn,draw,rectangle, rounded corners,minimum height=.5*\h,minimum width=0.9cm] (snn1) {SNN $\left(\theta_{1}^{(L)},\theta_{2}^{(L)}\right)$};
    \node[right=\xdist of snn,draw,circle, inner sep=0pt,radius=0.1cm] (a1) {$+$};
    \node[right=\xdist of a1,draw,circle, inner sep=0pt,radius=0.1cm] (m1) {$\times$};

    \draw[->] (L1) to (\tikztostart -| abs.west);
    \draw[->] (L2) to (\tikztostart -| abs.west);
    \draw[->] (L3) to (\tikztostart -| abs.west);

    \draw[->] (L1 -| abs.east)+(0cm,7.5mm) to (\tikztostart -| snn0.west);
    \draw[->] (L1 -| abs.east)+(0cm,6mm) to (\tikztostart -| snn0.west);
    \draw[->] (L1 -| abs.east)+(0cm,4.5mm) to (\tikztostart -| snn0.west);

    \draw[->] (L2 -| abs.east)+(0cm,3.5mm) to (\tikztostart -| snn.west);
    \draw[->] (L2 -| abs.east)+(0cm,2mm) to (\tikztostart -| snn.west);
    \draw[->] (L2 -| abs.east)+(0cm,.5mm) to (\tikztostart -| snn.west);

    \draw[->] (L3 -| abs.east)+(0cm,-2mm) to (\tikztostart -| snn1.west);
    \draw[->] (L3 -| abs.east)+(0cm,-3.5mm) to (\tikztostart -| snn1.west);
    \draw[->] (L3 -| abs.east)+(0cm,-5mm) to (\tikztostart -| snn1.west);

    \draw[->] (snn0) -| (a1);
    \draw[->] (snn1) -| (a1);
    \draw[->] (snn) -- (a1);
    \draw[->] (a1) -- (m1);

    \node[right=\xdist of L1,draw,circle,inner sep=1, fill=black] (b1) {};
    \node[right=\xdist+0.1cm of L2,draw,circle,inner sep=1, fill=black] (b2) {};,rotate=90,anchor=north
    \node[right=\xdist+0.13cm of L3,draw,circle,inner sep=1, fill=black] (b3) {};
    \node[below=-0.15cm of b2] () {$\vdots$};
    \node[below=.05cm of snn, text height=0.2cm] () {$\vdots$};

    \node[above=\ydist*0.5 of abs,draw, rectangle, rounded corners,minimum height=\h] (sgn) {sgn$(\cdot)$};
    \node[right=\xdist of sgn,draw,rectangle,rounded corners,,minimum height=\h,minimum width=0.9cm] (product) {$\prod$};

    \draw[->] (b1) -- +(0cm,2.1cm) to (\tikztostart -|sgn.west);
    \draw[->] (b2) |- (sgn);
    \draw[->] (b3) -- +(0cm,2.85cm) to (\tikztostart -|sgn.west);

    \draw[->] (sgn)++(0.55cm,0.15cm) to (\tikztostart -| product.west);
    \draw[->] (sgn) to (\tikztostart -| product.west);
    \draw[->] (sgn)++(0.55cm,-0.15cm) to (\tikztostart -| product.west);

    \draw[->] (product) -| (m1);

    \node[right=\xdist of m1,circle,draw,inner sep=2pt] (LI) {LI};
    \draw[->] (m1) -- (LI);
    \draw[->] (LI) -- ++(.7cm,0cm) node[pos=0.9,below] () {$L_{i \leftarrow j}^{[\mathrm{c}]}$};

\end{tikzpicture}
    }
    \caption{SCNU of the proposed \gls{MLE} decoder.}
    \label{fig:scnu_mle}
\end{figure}

To overcome this problem, we propose the \gls{MLE} decoder, which better approximates the \gls{NMS} decoder.
The \gls{MLE} decoder is derived from the \gls{ELENA} decoder by modifying its SCNU to allow for a higher range and resolution of the output of an SCNU. 
Fig.~\ref{fig:scnu_mle} shows the modified SCNU:
\gls{MLE} uses $L$ parallel \glspl{SNN} per SCNU, each \gls{SNN} having its own threshold $\theta_{1}^{(\ell)},\, \ell \in \{1,2,\ldots,L\}$ and amplitude $\theta_{2}^{(\ell)},\, \ell \in \{1,2,\ldots,L\}$.
The outputs of the \glspl{SNN} are combined by a summation:
\vspace{-2mm}
\begin{align}
    \alpha_{i\leftarrow j}^{[\mathrm{c}]} &\approx  \sum_{\ell=1}^L \theta_{2}^{(\ell)} \Theta\left( \min_{i' \in \mathcal{M}(j)\setminus \{i\}} \left| L_{i'\rightarrow j}^{[\mathrm{v}]} \right| - \theta_{1}^{(\ell)}  \right) \, .
    \vspace{-3mm}
\end{align}
Thus, an \gls{SNN} contributes to the summation and therefore to the output of the SCNU by $\theta_2^{(\ell)}$, if the smallest input message $| L_{i'\rightarrow j}^{[\mathrm{v}]} |$ exceeds the threshold $\theta_1^{(\ell)}$ of the respective \gls{SNN}. 
Hence, if the smallest input message to the SCNU has a small amplitude, a smaller number of \glspl{SNN} is triggered, and thus, an output message $| L_{i'\leftarrow j}^{[\mathrm{c}]}|$ with a small amplitude is created.
If the smallest input message has a large amplitude, more \glspl{SNN} contributes to the sum, creating an output message with a larger amplitude.
Depending on the chosen values of $\theta_1^{(\ell)}$, the smallest input message $\min_{i' \in \mathcal{M}(j)\setminus \{i\}} | L_{i'\rightarrow j}^{[\mathrm{v}]} |$ and therefore the uncertainty carried by this message can be resolved with finer granularity and in a larger range.
Depending on the chosen values of $\theta_2^{(\ell)}$, the CN update messages $L_{i\leftarrow j}^{[\mathrm{c}]}$ of the \gls{MLE} decoder have a higher resolution and larger dynamic range, too, increasing their accuracy.
Consequently, the resolution and dynamic range of the following VN update $L_{i'\rightarrow j}^{[\mathrm{v}]}$ is also increased.
Hence, by avoiding the all-or-nothing behavior of the SCNU of the \gls{ELENA} decoder, the \gls{MLE} decoder introduces a higher resolution and larger dynamic range for all LLRs.

\begin{figure}
    \centering
    \resizebox{.42\textwidth}{!}{\begin{tikzpicture}
    \begin{axis}[
        axis lines=middle,
        xlabel = {\small $\stackrel[{i' \in \mathcal{M}(j)\setminus \{i\}}]{}{\mathrm{min}} \left| L_{i'\rightarrow j}^{[\mathrm{v}]} \right|$},
        xlabel style={at={(ticklabel* cs:0.95)}, anchor=north},
        ylabel={\small $\left|L_{i\leftarrow j}^{[\mathrm{c}]}\right|$},%
        xtick = {0,1,2,3,4,5,6,7,8},
        xticklabels ={,,,,,$5\theta_1$,,,,},
        ytick = {0,1,2,3,4,5,6,7,8},
        yticklabels ={,,,,,$5\theta_2$,,,},
        ymin=-0.5, ymax=8.5,
        xmin=-0.5, xmax=8.5,
        domain=-4:4,
        samples=300,
        thick,
        legend style={at={(axis cs:0.4,7.0)},anchor=north west}
    ]

    \addplot[mark=none, black, very thick,] coordinates {(0,0) (9,9)};
    \addplot[mark=*, gray, very thick,] coordinates {(0,0) (1,0) (1,1) (2,1) (3,1) (4,1) (5,1) (6,1) (7,1) (8,1) (9,1)};
    \addplot[mark=none, KITorange, very thick,loosely dashdotdotted] coordinates {(0,0) (1,0) (1,1) (2,1) (2,2) (9,2)};
    \addplot[mark=none, KITcyanblue, very thick,dashed] coordinates {(0,0) (1,0) (1,1) (2,1) (2,2) (3,2) (3,3) (4,3) (4,4) (9,4)};
    \addplot[mark=none, KITpurple, very thick, dotted] coordinates {(0,0) (1,0) (1,1) (2,1) (2,2) (3,2) (3,3) (4,3) (4,4) (5,4) (5,5) (6,5) (6,6) (7,6) (7,7) (8,7) (8,8) (9,8)};

    \addlegendentry{\small NMS}
    \addlegendentry{\small$L=1$}
    \addlegendentry{\small$L=2$}
    \addlegendentry{\small$L=4$}
    \addlegendentry{\small$L=8$}

    \end{axis}
\end{tikzpicture}}
    \caption{Characteristic of an SCNU of the proposed \gls{MLE} decoder with $L$ parallel \glspl{SNN} and ${\theta_{1}^{(\ell)} = \ell \cdot \theta_{1},\, \ell \in \{1,2,\ldots,L\}},$ and $\theta_2^{(\ell)}=\theta_2$. For comparison, the characteristic of the \gls{NMS} decoder is given.}
    \label{fig:scnu_character}
\end{figure}
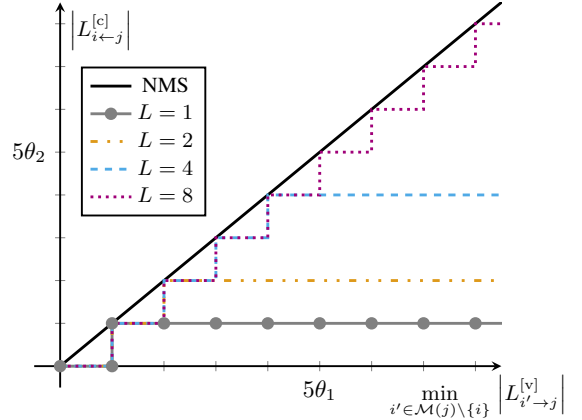

Inside an SCNU of the \gls{MLE} decoder, the threshold $\theta_1^{(\ell)}$ of the $\ell$-th \gls{SNN} is chosen as ${\theta_{1}^{(\ell)} = \ell \cdot \theta_{1},\, \ell \in \{1,2,\ldots,L\}}$, where $\theta_1$ denotes the spacing between neighbouring thresholds.
The output amplitude $\theta_2^{(\ell)}$ of each \gls{SNN} is shared among all \glspl{SNN} inside the SCNU, ${\theta_2^{(\ell)}=\theta_2, \, \ell \in \{1,2,\ldots,L\}}$. 
Hence, the sum of the outputs of all \glspl{SNN} is a multiple of $\theta_2$, resulting in ${L+1}$ discrete values $\alpha_{i\leftarrow j}^{[\mathrm{c}]}$ can have and ${2L+1}$ discrete values for $L_{i\leftarrow j}^{[\mathrm{c}]}$.
Fig.~\ref{fig:scnu_character} shows the characteristics of an SCNU for different numbers $L$ of parallel \glspl{SNN} with fixed $\theta_1$ and $\theta_2$.
If $L$ is increased, the range which avoids clipping of the smallest input amplitude is increased by the factor $L$.
Furthermore, the range of the output message is increased by a factor of $L$, too. 

To further reduce the dimensionality of the search space, we couple $\theta_2=\gamma\cdot \theta_1,\, \gamma \in \mathbb{R}$, and apply a line search over~$\theta_1$ w.r.t. the BER.

\section{Results and Discussion}
We implemented and evaluated the performance of the proposed \gls{MLE} decoder for two different \gls{LDPC} codes:
an $(38400,\,30720)$ regular \gls{LDPC} code and an \gls{FG} $(273,\,191)$ regular \gls{LDPC} code.
The latter was already used in~\cite{ELENA}.
The $(38400,\,30720)$ code has ${d_\mathrm{v}=3}$ and ${d_\mathrm{c}=15}$. It was constructed by adding columns with `1`s at random positions, where the number of `1`s in a row is below $d_\mathrm{c}$.
To prevent 4-cycles, only columns that do not create 4-cycles were added~\cite{RS01}.
If no column matches the constraints, the process is reset to a random previous column. 
The $(273,\,191)$ code has $d_\mathrm{v}=17$ and $d_\mathrm{c}=17$, and the parity-check matrix is highly rank deficient.

After transmission over an AWGN channel, $y_i$ represents the received value corresponding to the $i$-th code bit of a codeword.
The initial bit-wise LLRs are given by $L_i=y_i\cdot L_\mathrm{c}$, where $L_\mathrm{c}$ is the channel reliability parameter. 
Typically, for an AWGN channel, $L_\mathrm{c}=4\frac{E_\mathrm{s}}{N_0}=\frac{2}{\sigma^2}$. 
Hence, $L_\mathrm{c}$ needs to be adjusted according to the SNR.
However, depending on the code, the \gls{MLE} decoder is optimized at a fixed $\frac{E_\mathrm{s}}{N_0}$; this value is used during evaluation regardless of the actual~SNR.

We compare the performance of the \gls{MLE} decoder against the \gls{ELENA} decoder, as well as the \gls{SPA} decoder, the \gls{MS} decoder, and the \gls{NMS} decoder.
All decoders use 20 decoding iterations, which offers a reasonable trade-off between decoding convergence and computation time.
Like the \gls{MLE} decoder, the \gls{ELENA} decoder is optimized and evaluated at a fixed $L_\mathrm{c}$. In contrast, for all other benchmark decoders $L_\mathrm{c}$ is adjusted to match the actual~$E_\mathrm{s}/N_0$.

\subsection{Parameters} 
As in~\cite{ELENA}, for both the \gls{MLE} and \gls{ELENA} decoder, the parameters of all spiking \gls{LIF} neurons are chosen to $\tau_\mathrm{s}=\SI{1}{\milli\second}$, $\tau_\mathrm{m}=\SI{1}{\milli\second}$ and $v_\mathrm{th}=1$.
In~\cite[Fig.~6a]{ELENA}, the parameters $\tau_\mathrm{s}=\SI{1}{\milli\second}$ and $\tau_\mathrm{m}=\SI{1}{\milli\second}$ for the \gls{LI} neuron of the $(273,\,191)$ code and \gls{ELENA} decoder yield the best decoder performance.
Since a parameter search did not yield better parameters for both decoders and codes, the parameters of the \gls{LI} neuron are fixed to~${\tau_\mathrm{s}=\SI{1}{\milli\second}}$ and~${\tau_\mathrm{m}=\SI{1}{\milli\second}}$.

For the $(38400,\,30720)$ code, the \gls{MLE} decoder was implemented with $L=4$, $L=8$, and $L=16$ parallel \glspl{SNN} per SCNU.
The resulting decoders were optimized for $L_\mathrm{c}=\SI{1.0}{\decibel}$, $L_\mathrm{c}=\SI{2.8}{\decibel}$, and $L_\mathrm{c}=\SI{3.0}{\decibel}$. 
For the $(273,\,191)$ code, only $L=16$ with $L_\mathrm{c}=\SI{3.0}{\decibel}$ was chosen. 
Tab.~\ref{tab:param_results} shows an overview of the decoders displayed in this paper, where the term \mbox{``$L$-MLE-$L_\mathrm{c}$''} denotes the MLE decoder with $L$ parallel SNNs per SCNU, optimized at a fixed channel SNR and channel reliability parameter of $L_\mathrm{c}$. Hence, the \mbox{$8$-MLE-$2.8$} is the decoder with $L=8$ \glspl{SNN} optimized at $L_\mathrm{c}=\SI{2.8}{\decibel}$. 
To show the effect of adapting $L_\mathrm{c}$, for the $(38400,\,30720)$ code and $L=16$, we supply a decoder that matches its $L_\mathrm{c}$ according to the channel SNR ($16L$-MLE-$L_\mathrm{c}$).
Since the novel decoder only slightly improves the performance for the $(273,\,191)$ code, only the \gls{ELENA} decoder ($1$-MLE-$3.5$) and the \mbox{$16$-MLE-$3.0$} are implemented.

\begin{figure}
    \centering
    \resizebox{0.42\textwidth}{!}{\pgfplotsset{
layers/my layer set/.define layer set={
background,
main,
up
}{
 },
    set layers=my layer set,
}

\begin{tikzpicture}[spy using outlines={rectangle, magnification=2.7, size=2cm, connect spies}]
    \def\lwidth{1.5}
    \def\opac{100}
    \def\marksz{1.5pt}
    \def\markszl{3pt}

    \def\spywidth{2.0cm}
    \def\spyheigth{1.5cm}

    \begin{axis}[
        ymode = log,
        xscale=1,
        xlabel style = {align=center},
        xlabel=$\theta_1$,
        x label style={at={(axis description cs:0.5,0.02)},anchor=north},
        ylabel=BER,
        y label style={at={(axis description cs:0.00,0.4)},anchor=west},
        xtick = {1.0,2.0,3.0,4.0},
        xticklabels={$1.0$,$2.0$,$3.0$,$4.0$},
        grid=major,
        legend cell align={left},
        legend style={
            at={(0.64,0.00)},
            anchor=south west,
            fill opacity = 1,
            draw opacity = 1, 
            text opacity = 1,
            nodes={scale=0.9, transform shape}
        },
        xmin=0.1,xmax=4.0,
        ymin=5e-6,ymax=1e-1,
        axis line style=thick,
        tick label style={/pgf/number format/fixed},
        legend image post style={scale=0.9},
        ]

        \addplot[mark=none, color=gray,line width=\lwidth, opacity= \opac] table [x = TH, y = Elena]{./figures/Threshold_Sweep.txt};
        \addplot[dotted,mark=none, color=KITpurple,line width=\lwidth, opacity= \opac] table [x = TH, y = MLElena4]{./figures/Threshold_Sweep.txt};
        \addplot[dashed,mark=none, color=KITorange,line width=\lwidth, opacity= \opac] table [x = TH, y = MLElena8_10]{./figures/Threshold_Sweep.txt};
        \addplot[dashed,mark=none, color=KITpurple,line width=\lwidth, opacity= \opac] table [x = TH, y = MLElena8_28]{./figures/Threshold_Sweep.txt};
        \addplot[dashed,mark=none, color=KITcyanblue,line width=\lwidth, opacity= \opac] table [x = TH, y = MLElena8_30]{./figures/Threshold_Sweep.txt};
        \addplot[solid,mark=none, color=KITpurple,line width=\lwidth, opacity= \opac] table [x = TH, y = MLElena16]{./figures/Threshold_Sweep.txt};

        \addlegendimage{} \addlegendentry{\small $1$-MLE-$2.8$}
        \addlegendimage{} \addlegendentry{\small $4$-MLE-$2.8$}
        \addlegendimage{} \addlegendentry{\small $8$-MLE-$1.0$}
        \addlegendimage{} \addlegendentry{\small $8$-MLE-$2.8$}
        \addlegendimage{} \addlegendentry{\small $8$-MLE-$3.0$}
        \addlegendimage{} \addlegendentry{\small $16$-MLE-$2.8$}

    \begin{pgfonlayer}{up}
        
    \end{pgfonlayer}
\end{axis}
\end{tikzpicture}}
    \vspace{-2mm}
    \caption{BER curve for the $(38400,\,30720)$ regular \gls{LDPC} code as a function of $\theta_1$, where $\theta_2$ is chosen to be equal to $\theta_1$, i.e., $\theta_2 = \theta_1$.}
    \label{fig:Sweep}
\end{figure}
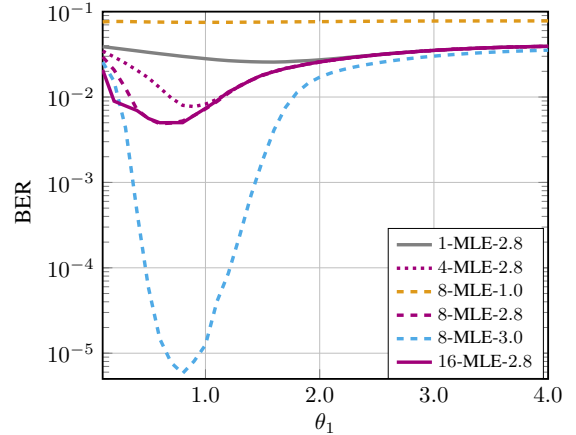

The coupling factor $\gamma$ of Sec.~\ref{sec:ML-ELENA} is chosen experimentally.
For the $(38400,\,30720)$ code and both the \gls{ELENA} and \gls{MLE} decoder, we choose $\gamma=1$, hence $\theta_2 = \theta_1$.
For the $(273,\,191)$ code and the MLE decoder with $L=16$, we choose $\gamma=0.5$, while the parameters of the \gls{ELENA} decoder are obtained from~\cite[Fig.~6]{ELENA}.
Fig.~\ref{fig:Sweep} shows the BER over $\theta_1$ for varying numbers of levels~$L$. With increasing $L$, the performance improves. 
The parameters $\theta_1$ and $\theta_2$, depending on the code and architecture of the decoder, are summarized in Tab.~\ref{tab:param_results}.
Furthermore, Fig.~\ref{fig:Sweep} shows the BER over $\theta_1$ for different $E_\mathrm{b}/N_0$.
Three sweeps are conducted at $E_\mathrm{b}/N_0$ values of $\SI{1.0}{\decibel}$ ($8$-MLE-$1.0$), $\SI{2.8}{\decibel}$ ($8$-MLE-$2.8$), and $\SI{3.0}{\decibel}$ ($8$-MLE-$3.0$) for the \gls{MLE} decoder with $L=8$ steps.
Changing the operation point affects the position of the minimum BER and broadens the region of suitable $\theta_1$.

\begin{table}
    \centering
    \begin{tabular}{cl}
        \toprule
          Code & \hspace{1.4mm} Decoder \hspace{9mm}   $\theta_1$ \hspace{4mm}    $\theta_2$  \\
          \midrule
         $(38400,\,30720)$ &  \begin{tabular}{lll}
         $1$-MLE-$2.8$ & 1.5\textcolor{white}{0} & 1.5 \\
         $4$-MLE-$2.8$ & 0.9 & 0.9 \\
         $8$-MLE-$1.0$ & 0.7 & 0.7 \\
         $8$-MLE-$2.8$ & 0.7 & 0.7 \\
         $8$-MLE-$3.0$ & 0.8 & 0.8 \\
         $16$-MLE-$2.8$& 0.7 & 0.7 \\
         $16$-MLE$~L_\mathrm{c}$ & 0.7 & 0.7 \\
         \end{tabular} \\
         \midrule
         $(273,\,191)$ & \begin{tabular}{lll}
         $1$-MLE-$3.5$ & 2.0 & 1.4 \\
         $16$-MLE-$3.0$& 0.95 & 0.475 \\
        \end{tabular} \\
        \bottomrule
    \end{tabular}
    \caption{$\theta_1$ and $\theta_2$ of the SNN-based decoders for the $(38400,\,30720)$ regular \gls{LDPC} code and the \gls{FG} $(273,\,191)$ regular \gls{LDPC} code.}
    \label{tab:param_results}
\end{table}
    
\subsection{Results}
\begin{figure}
    \centering
    \begin{subfigure}[b]{.48\textwidth}
        \centering
        \resizebox{\textwidth}{!}{\pgfplotsset{
layers/my layer set/.define layer set={
background,
main,
up
}{
 },
    set layers=my layer set,
}

\begin{tikzpicture}[spy using outlines={rectangle, magnification=2.7, size=2cm, connect spies}]
    \def\lwidth{1.5}
    \def\opac{100}
    \def\marksz{1.5pt}
    \def\markszl{3pt}

    \def\spywidth{2.0cm}
    \def\spyheigth{1.5cm}

    \begin{axis}[
        ymode = log,
        xscale=1,
        xlabel style = {align=center},
        xlabel=$E_\mathrm{b}/N_0$ (dB),
        x label style={at={(axis description cs:0.5,0.02)},anchor=north},
        ylabel=BER,
        y label style={at={(axis description cs:0.00,0.4)},anchor=west},
        xtick = {2.5,3.0,3.5},
        xticklabels={$2.5$,$3.0$,$3.5$},
        grid=major,
        legend cell align={left},
        legend style={
            at={(0.02,0.02)},
            anchor=south west,
            fill opacity = .8,
            draw opacity = 1, 
            text opacity = 1,
            nodes={scale=1, transform shape}
        },
        xmin=2.5,xmax=3.5,
        ymin=4e-9,ymax=1e-1,
        axis line style=thick,
        tick label style={/pgf/number format/fixed},
        legend image post style={scale=0.9},
        ]

        \coordinate (spypoint) at (axis cs:3.05,1.0e-6); %
        \coordinate (spyviewer) at (axis cs:4.1,1.0e-7); %
        \draw [fill=white] ($(spyviewer)+(1.2cm,0.7cm)$) rectangle ($(spyviewer)-(1.2cm,0.7cm)$);

        \addplot[mark=none, color=black,line width=\lwidth, opacity= \opac] table [x = ebn0, y = spa]{./figures/38400-30720_20Iter.txt};

        \addplot[dashed,color=KITgreen,mark=square*,every mark/.append style={solid,fill=KITgreen},mark size=\marksz,line width=\lwidth, opacity =\opac] table [x=ebn0, y=ms]{./figures/38400-30720_20Iter.txt};
        
        \addplot[dotted, mark=*,mark size =\marksz,every mark/.append style={solid, fill=KITcyanblue}, color=KITcyanblue,line width=\lwidth, opacity =\opac] table [x = ebn0, y =nms]{./figures/38400-30720_20Iter.txt};

        \addplot[dotted, mark=diamond*,mark size =\marksz,every mark/.append style={solid, fill=gray}, color=gray,line width=\lwidth, opacity =\opac] table [x = ebn0, y = elena]{./figures/38400-30720_20Iter.txt};
        
        \addplot[dotted,color=KITpurple,line width=\lwidth, opacity =\opac] table [x = ebn0, y = mlelena4]{./figures/38400-30720_20Iter.txt};

        \addplot[dashed,color=KITorange, line width=\lwidth] table [x=ebn0, y=mlelena8_10]{./figures/38400-30720_20Iter.txt};
        \addplot[dashed,color=KITpurple, line width=\lwidth] table [x=ebn0, y=mlelena8_28]{./figures/38400-30720_20Iter.txt};
        \addplot[dashed,mark=none, color=KITcyanblue, line width=\lwidth] table [x=ebn0, y=mlelena8_30]{./figures/38400-30720_20Iter.txt};
        
        \addplot[solid,color=KITpurple, line width=\lwidth] table [x=ebn0, y=mlelena16]{./figures/38400-30720_20Iter.txt};
        \addplot[solid,color=kit-purple50, line width=\lwidth] table [x=ebn0, y=mlelena16nflc]{./figures/38400-30720_20Iter.txt};
        
        \addlegendimage{} \addlegendentry{\small SPA}
        \addlegendimage{} \addlegendentry{\small MS}
        \addlegendimage{} \addlegendentry{\small NMS}
        \addlegendimage{} \addlegendentry{\small $1$-MLE-$2.8$}%
        \addlegendimage{} \addlegendentry{\small $4$-MLE-$2.8$}%
        \addlegendimage{} \addlegendentry{\small $8$-MLE-$1.0$}%
        \addlegendimage{} \addlegendentry{\small $8$-MLE-$2.8$}%
        \addlegendimage{} \addlegendentry{\small $8$-MLE-$3.0$}%
        \addlegendimage{} \addlegendentry{\small $16$-MLE-$2.8$}%
        \addlegendimage{} \addlegendentry{\small $16$-MLE-$L_\mathrm{c}$}%

    \begin{pgfonlayer}{up}
        
    \end{pgfonlayer}
\end{axis}
\end{tikzpicture}}
        \vspace{-5mm}
        \caption{BER curve for the $(38400,\,30720)$ regular \gls{LDPC} code.}
        \label{fig:res_38400}
    \end{subfigure}\\
    \vspace{3mm}
    \begin{subfigure}[b]{.48\textwidth}
        \centering
        \resizebox{\textwidth}{!}{\pgfplotsset{
layers/my layer set/.define layer set={
background,
main,
up
}{
 },
    set layers=my layer set,
}

\begin{tikzpicture}[spy using outlines={rectangle, magnification=2.7, size=2cm, connect spies}]
    \def\lwidth{1.5}
    \def\opac{100}
    \def\marksz{1.5pt}
    \def\markszl{3pt}

    \def\spywidth{2.0cm}
    \def\spyheigth{1.5cm}

    \begin{axis}[
        ymode = log,
        xscale=1,
        xlabel style = {align=center},
        xlabel=$E_\mathrm{b}/N_0$ (dB),
        x label style={at={(axis description cs:0.5,0.02)},anchor=north},
        ylabel=BER,
        y label style={at={(axis description cs:0.00,0.4)},anchor=west},
        xtick = {0.5,1.0,1.5,2.0,2.5,3.0,3.5,4.0,4.5,5.0},
        xticklabels={$0.5$,$1.0$,$1.5$,$2.0$,$2.5$,$3.0$,$3.5$,$4.0$,$4.5$,$5.0$},
        grid=major,
        legend cell align={left},
        legend style={
            at={(0.02,0.01)},
            anchor=south west,
            fill opacity = 0.8,
            draw opacity = 1, 
            text opacity = 1,
        },
        xmin=0.5,xmax=5.0,
        ymin=8e-8,ymax=1,
        axis line style=thick,
        tick label style={/pgf/number format/fixed},
        legend image post style={scale=0.9},
        ]

        \coordinate (spypoint) at (axis cs:4.0,1.0e-6); %
        \coordinate (spyviewer) at (axis cs:4.1,1e-1); %
        \draw [fill=white] ($(spyviewer)+(1.2cm,0.7cm)$) rectangle ($(spyviewer)-(1.2cm,0.7cm)$);
        
        \spy[width=2.4cm,height=1.4cm, every spy on node/.append style={ultra thin},thin,line width=0.01, spy connection path={
        \draw [opacity=0.5] (tikzspyonnode.south west) -- (tikzspyinnode.south west);
        \draw [opacity=0.5] (tikzspyonnode.south east) -- (tikzspyinnode.south east);
        \draw [opacity=0.5] (tikzspyonnode.north west) -- (intersection of tikzspyinnode.north west--tikzspyonnode.north west and tikzspyinnode.south east--tikzspyinnode.south west);
        \draw [opacity=0.5] (tikzspyonnode.north east) -- (intersection of  tikzspyinnode.north east--tikzspyonnode.north east and tikzspyinnode.south east--tikzspyinnode.south west);
        ;}] on (spypoint) in node at (spyviewer); %

        \addplot[mark=none, color=black,line width=\lwidth, opacity= \opac] table [x = ebn0, y = spa]{./figures/FG_273-191_20Iter.txt};
        
        \addplot[dashed,mark=square*,mark size =\marksz,mark repeat=5,every mark/.append style={solid, fill=KITgreen},color=KITgreen,line width=\lwidth, opacity =\opac] table [x=ebn0, y=ms]{./figures/FG_273-191_20Iter.txt};
        
        \addplot[dotted, mark=*,mark size =\marksz,every mark/.append style={solid, fill=KITcyanblue},mark repeat=5, color=KITcyanblue,line width=\lwidth, opacity =\opac] table [x = ebn0, y =nms]{./figures/FG_273-191_20Iter.txt};

        \addplot[dotted, mark=diamond*,mark repeat=5,mark size =\marksz,every mark/.append style={solid, fill=KITorange}, color=KITorange,line width=\lwidth, opacity =\opac] table [x = ebn0, y = elena]{./figures/FG_273-191_20Iter.txt};
        
        \addplot[solid,color=KITpurple, line width=\lwidth] table [x=ebn0, y=mlelena16]{./figures/FG_273-191_20Iter.txt};

        \addplot[dotted, mark=*,mark size =\marksz,every mark/.append style={solid, fill=KITcyanblue},mark repeat=5, color=KITcyanblue,line width=\lwidth, opacity =\opac] table [x = ebn0, y =nms]{./figures/FG_273-191_20Iter.txt};
        
        \addlegendimage{} \addlegendentry{\small SPA}
        \addlegendimage{} \addlegendentry{\small MS}
        \addlegendimage{} \addlegendentry{\small NMS}
        \addlegendimage{} \addlegendentry{\small $1$-MLE-$3.5$}%
        \addlegendimage{} \addlegendentry{\small $16$-MLE-$3.0$}%

    \begin{pgfonlayer}{up}
        
    \end{pgfonlayer}
\end{axis}
\end{tikzpicture}}
        \vspace{-5mm}
        \caption{BER curve for the FG ($273$,$191$) LDPC code.}
        \label{fig:res_273}
    \end{subfigure}
    \vspace{2mm}
    \caption{BER curve of the \gls{MLE} decoder and reference decoders.}
    \label{fig:results}
\end{figure}
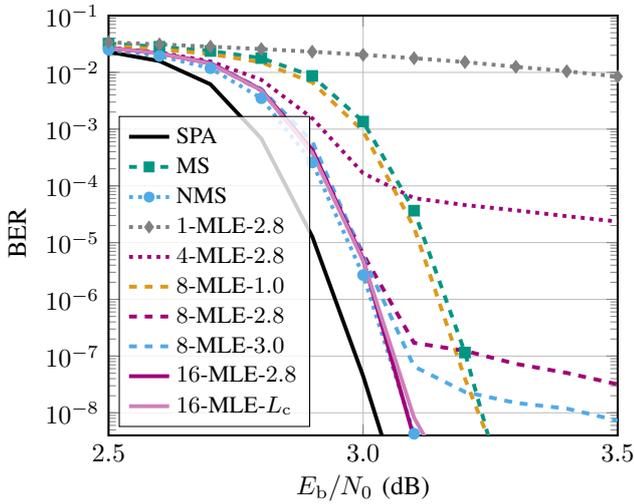
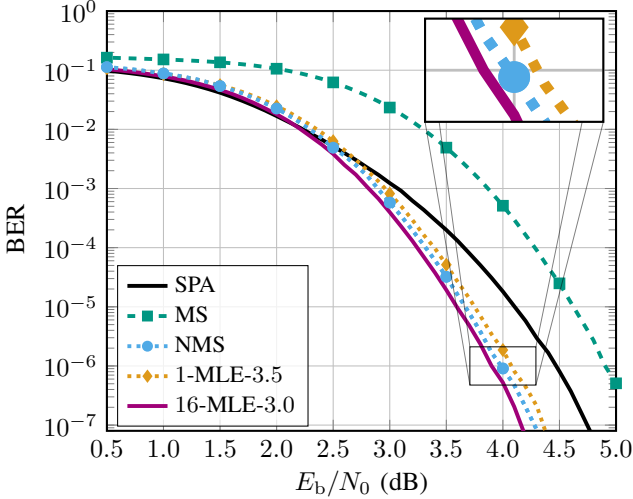

\vspace*{-1mm}
Fig.~\ref{fig:results}(a) shows the BER results for the $(38400,\,30720)$ regular \gls{LDPC} code. 
Due to its small VN degree ${d_\mathrm{v}=3}$, the previously proposed \gls{ELENA} decoder is not competitive.
However, with an increasing number of levels $L$, the performance of the \gls{MLE} decoder improves, nearly reaching the performance of the \gls{NMS} decoder.
With four levels, the \gls{MLE} decoder performs close to the \gls{NMS} decoder up to an $E_\mathrm{b}/N_0$ of $\SI{2.7}{\decibel}$.
We attribute the error floor above $\SI{2.7}{\decibel}$ dB to an insufficient number of threshold levels and, hence, a low dynamic range.

Fig.~\ref{fig:results}(a) also shows how the operation point alters the performance in terms of BER over $E_\mathrm{b}/N_0$.
The $8$-MLE-$2.8$ decoder performs slightly better than the $8$-MLE-$3.0$ decoder in the $E_\mathrm{b}/N_0$ region of $\SI{2.5}{\decibel}$ to $\SI{3.0}{\decibel}$, but has an higher error floor at an $\mathrm{BER}$ of $2\cdot 10^{-7}$ compared to at an $\mathrm{BER}$ of $8\cdot10^{-8}$. 
Fig.~\ref{fig:res_38400} shows that the decoder with fixed channel reliability, \mbox{$16$-MLE-$3.1$}, performs as the decoder with adapting channel reliability,~\mbox{$16$-MLE-$L_\mathrm{c}$}.

Fig.~\ref{fig:results}(b) shows the BER results for the \gls{FG} $(273,\,191)$ regular \gls{LDPC} code.
The $16$-MLE-$3.5$ decoder and the \gls{ELENA} ($1$-MLE-$3.5$) decoder achieve similar performance, with the $16$-MLE-$3.5$ decoder slightly outperforming the \gls{ELENA} decoder. Both decoders outperform the \gls{MS} and \gls{SPA} decoders.

\subsection{Discussion}
For the $(38400,\,30720)$ code, we have shown that the \gls{ELENA} decoder is not competitive, which is due to the small VN degree $d_\mathrm{v}$ and, thus, low dynamic range in the \gls{ELENA} decoder.
By introducing multiple parallel \glspl{SNN} with distinct threshold levels $\theta_1^{(\ell)}$ into the decoding process, the extended decoder succeeds for codes with small $d_\mathrm{v}$.  
With an increasing number of \glspl{SNN}, the dynamic range of the variable-to-check-node-messages is increased, improving the decoding performance. 
If the VN degree $d_\mathrm{v}$ of the code is sufficiently large, e.g., for the $(273,\,191)$ code, the \gls{ELENA} decoder already achieves a performance similar to the NMS decoder. 
Hence, introducing multiple threshold levels does not increase performance significantly. 

The \gls{MLE} provides a trade-off between performance and complexity.
With an increasing number of threshold levels $L$, the performance of the decoder improves at the cost of increasing complexity.
Thus, the \gls{MLE} decoder enables the decoding of \gls{LDPC} codes with small VN degree~$d_\mathrm{v}$.
Future work will investigate the parameter space $\theta_1^{(\ell)}$, $\theta_2^{(\ell)}$, $\tau_\mathrm{s}$ and $\tau_\mathrm{m}$ more thoroughly.

\section{Conclusion}
In this work, we introduced the \gls{MLE} decoder, an extension of the \gls{ELENA} decoder.
The \gls{MLE} decoder approximates the CN update equation using \glspl{SNN} in a SCNU.
However, compared to the \gls{ELENA} decoder, the \gls{MLE} decoder enhances the resolution and dynamic range of the variable-to-check-node messages, enabling successful decoding of regular \gls{LDPC} codes with a small VN degree $d_\mathrm{v}$.
The proposed decoder consists of multiple parallel SNNs inside an SCNU, each SNN having its own threshold and output amplitude. 
For each SNN inside the SCNU, the threshold is increased by $\theta_1$, where the output amplitude $\theta_2$ is shared among all \glspl{SNN}.

While the \gls{ELENA} decoder performs poorly for a (38400, 30720) regular LDPC code, the proposed \gls{MLE} decoder achieves performance close to the \gls{NMS} decoder.
For the  FG (273, 191) LDPC code, the proposed decoder also achieves similar performance as the \gls{NMS} decoder.
Furthermore, similar to the \gls{ELENA} decoder, the \gls{MLE} decoder is optimized at a single $E_\mathrm{b}/N_0$ and generalizes well over the whole $E_\mathrm{b}/N_0$ range. 
We conclude that the \gls{MLE} decoder enables successful decoding of regular \gls{LDPC} codes with small $d_\mathrm{v}$.

\end{document}